\documentclass[12pt]{article}

\author{Andrei Yu. Khrennikov\\
Center for Mathematical
Modelling \\ in Physics and Cognitive Sciences,\\
University of V\"axj\"o, S-35195, Sweden\\
Email:Andrei.Khrennikov@msi.vxu.se}

\title{The Brain as Quantum-like Machine Operating on
Subcognitive and Cognitive Time Scales}

\begin{document}

\maketitle

\begin{abstract}
We propose a {\it quantum-like} (QL) model of the functioning of
the brain. It should be sharply distinguished from the
reductionist {\it quantum} model. By the latter cognition is
created by {\it physical quantum processes} in the brain. The
crucial point of our modelling is that discovery of the
mathematical formalism of quantum mechanics (QM) was in fact
discovery of a very general formalism describing {\it consistent
processing of incomplete information} about contexts (physical,
mental, economic, social). The brain is an advanced device which
developed the ability to create a QL representation of contexts.
Therefore its functioning can also be described by the
mathematical formalism of QM. The possibility of such a
description has nothing to do with composing of the brain of
quantum systems (photons, electrons, protons,...). Moreover, we
shall propose a model in that the QL representation is based on
conventional neurophysiological model of the functioning of the
brain. The brain uses the QL rule (given by von Neumann trace
formula) for calculation of {\it approximative averages} for
mental functions, but the physical basis of mental functions is
given by neural networks in the brain. The QL representation has a
{\it temporal basis.}  Any cognitive process is based on (at
least) two time scales:  subcognitive time scale (which is very
fine) and cognitive time scale (which is essentially coarser).
\end{abstract}

\section{Introduction}

We start with formulation of a number of questions which will be considered in this paper. We shall try to find proper
answers to those questions. From the very beginning we  recognize that the present level of
experimental research in quantum physics, neurophysiology, cognitive science, and psychology is not high
enough to support (or to reject) our quantum-like (QL) model of cognition.

Although basic theoretical tools
behind our model are mathematics and physics, we would like to address this paper to  really multi-disciplinary
auditorium: philosophers, neurophysiologists, psychologists, mathematicians, physicists, people working in
cognitive and social sciences, economy (our model can be easily generalized from a single
brain to a collection of brains),
and even abnormal phenomena. Therefore we practically excluded mathematical technique from this paper. A
reader can find corresponding details in references on author's papers, e.g.,
Khrennikov (2005a,b, 2006b--d).

This paper has a detailed introduction which is in principle sufficiently complete to create a general
picture of our QL cognitive model. The last part of this paper is more complicated.
It can be interesting  for  neurophysiologists and psychologists, since
we couple our QL model with the temporal structure of processes in the functioning of the
brain. A number of time scales
will be discussed as possible candidates for QL and ``sub-QL'' (vs. cognitive and subcognitive) time scales.

\subsection{Questions and answers}

\medskip

1). Should one distinguish the mathematical formalism of QM from quantum physics?
Yes.

\medskip

2). Can be this formalism be applied outside physics, e.g. in cognitive science or sociology?
Yes. But corresponding QL models should be tested experimentally.

\medskip

3). Can one escape in the QL framework difficulties of existing ``really quantum'' models of cognition
(based on reduction of cognition to quantum physical processes in the brain)? Yes.

\medskip

4). Can  the present neurophysiological model of the functioning of the brain be combined with
a QL model? Yes, see also question 9.

\medskip

5). Can quantum formalism coexist with hidden variables?
It seems -- yes, but the problem is extremely complicated. Further theoretical and
experimental investigations should be performed.

\medskip

6). Did Bell's arguments (and corresponding experiments, e.g., Aspect et al (1982) and Weihs et al (1998),
Weihs (2007)) as well
as other ``NO-GO'' theorems (e.g. von Neumann, Kochen-Specker)
really imply that local realism is incompatible with the formalism of QM?
Not at all, but the problem is extremely complicated. Further theoretical and
experimental investigations should be performed.

\medskip

7). What (Who?) is quantum (QL): the consciousness or the unconsciousness?
By our model the consciousness is QL and the unconsciousness is classical
(the latter operates via classical neural networks). We remark that our viewpoint
is {\it opposite} to the conventional viewpoint of quantum reductionist approach. By
the latter the unconsciousness is quantum and the consciousness is classical.

\medskip

8). Are QL cognitive models consistent with ideas of Leibniz, Freud, Nietzsche, see also N\"a\"at\"anen (1992)
for modern neurophysioloical considerations of this problem, that consciousness
is a projection of huge ocean of unconscious information? Yes, by our interpretation
the mathematical formalism of QM describes {\it operation with flows of incomplete
information.}

\medskip

9). What is a neural mechanism of the QL conscious representation of information?
We propose a model in that the QL representation is based on the presence of variety
of time scales in processing of mental information in the brain.

\subsection{Short description of quantum-like model of cognition}

UN). {\it Unconsciousness.} Each brain operates with huge amounts of unconscious information
which is produced via neural activity. Some basic cognitive processes (at low levels of
mental organization) are performed in this unconscious neural representation.

PR). {\it Projection.}  The brain projects this unconscious information ocean on the QL representation. It is
a probabilistic representation. Information about neural activity is represented by a complex probability
amplitude, {\it  a mental wave function.}

CON). {\it Consciousness.} We identify such a QL projection with the {\it conscious representation.}
The consciousness operates with complex probability amplitudes. Its decision making is based on probabilities
and averages described by the mathematical formalism of QM.

COM). {\it The brain as QL computer.} Hence,  the consciousness operates  with algorithms  which can be described by
the mathematical formalism of QM. These are so called quantum algorithms.
Since the brain realizes them without appealing to ``really quantum physical processes in microworld,''
we can consider it as QL (and not as quantum) computer, cf. Penrose's model of the brain as
quantum computer.

HV). {\it Hidden mental variables.} Our QL model of cognition has
hidden variables -- states of neurons. Thus advantages of QL
computation, in contrast to really quantum computation, are not
due to superposition of states for individual systems (in our case
neurons), but due to parallel working of billions of neurons.  The
consciousness does not operate  in the neural representation. It
operates in the QL representation. Therefore it proceeds
essentially faster than classical computer\footnote{In particular,
I completely agree with Penrose's critique of the famous in
50s-80s project of artificial intelligence which was an attempt to
explain cognition in the framework of classical computations.}.
However, the QL representation is created by activity of a huge
number of neurons working parallel . This is a purely classical
explanation of the process of QL computation.

G). {\it The temporal origin of the QL representation.} We suppose that any conscious QL representation (and the
brain can operate with a number of such representations) is based on two time scales: sub-QL (subcognitive)
and QL (cognitive). Probabilities and averages with respect to classical stochastic processes for
pre-QL time scale are represented in the QL way.  Then consciousness operates with such QL probabilistic
objects on the QL (cognitive) time scale.

E).  The QL representation of averages is based on approximation of classical averages by using
the Taylor approximation of psychological functions. Operation with approximations (instead
of complete classical averages) tremendously increases the
speed of processing of probabilistic information.

\subsection{On the level of mathematics and physics used in this paper}

We remark that the only mathematical things which will be used in the last (more technical) part are
three formulas, two from probability theory (one from classical and another from quantum)
and third from mathematical analysis:

a) The classical average (also called mean value or mathematical expectation) of a random variable
is given by integral. In the case of a discrete random variable it is simply a normalized sum.

b) The quantum average is given by the operator-trace in Hilbert space.\footnote{In the case of discrete spectrum the operator trace is simply the sum
of diagonal elements of the corresponding matrix.}

c) Taylor's formula: any smooth function can be approximated by using its derivatives
(by  Taylor's polynomial).

However, one can proceed rather far (in any event through an extended introduction) before she will meet
mathematics at all. Even in the last part we escaped long mathematical formulas. There will be used
just a few mathematical symbols.

Regarding quantum physics we only assume that a reader is familiar with  such fundamental
problems of quantum foundations as Eistein-Polodosky-Rosen paradox, completeness/incompleteness
of QM, Bell's inequality, ``death of reality'', quantum nonlocality. One need
not know mathematical and physical details, but just the general picture of 80 years of debates
on quantum foundations.

\subsection{Quantum vs. quantum-like cognitive models}

The idea that the description of brain functioning, cognition, and
consciousness could not be reduced to the theory of neural networks
and dynamical systems (cf. Ashby (1952), Hopfield (1982), Amit
(1989), Bechtel and Abrahamsen (1991), Strogatz (1994),  van Gelder
(1995), van Gelder and Port (1995), Eliasmith (1996)) and that
quantum theory may play an important role in such a description has
been discussed in a huge variety of forms, see e.g. Whitehead (1929,
1933, 1939), Orlov (1982),
 Healey (1984), Albert and Loewer (1988, 1992),
 Lockwood (1989, 1996), Penrose (1989, 1994),
 Donald (1990, 1995, 1996), Jibu and Yasue (1992, 1994),
 Bohm and Hiley (1993), Stapp (1993), Aerts, D. and Aerts, S. (1995, 2007),
 Hameroff (1994, 1998), Loewer (1996), Hiley and Pylkk\"anen (1997),
 Deutsch (1997),  Barrett (1999),
 Khrennikov (1999a, 2000, 2002b, 2003b, 2004a, 2006a), Hiley (2000), Vitiello
 (2001), Choustova (2001, 2004), Behera et al (2005),
 Haven (2006),  Conte et al. (2007), Atmanspacher (2007)  and literature thereby.

This idea that QM might have some consequences for
cognitive science and psychology was discussed at many occasions
already by fathers of quantum theory. We can mention, for example,
attempts of Niels Bohr to apply the quantum principle of
complementarity to psychology (see A. Plotnitsky 2001, 2002, 2007
for discussions). We can also mention the correspondence between
Pauli and Jung about analogy between quantum and mental processes.

During the last 30 years it was done  a lot for the realization of
the very ambitious program of quantum reductionism. There were
various attempts to reduce mental processes to quantum physical
processes in the brain. Here we point out to fundamental works
Hameroff (1994, 1998) and Penrose (1989, 1994, 2005).

However, the quantum formalism provides essentially more
possibilities for modelling of physical, biological, and social
processes. One should distinguish QM as physical theory and its
formalism. In principle, there is nothing surprising that a
formalism which was originally developed for serving for one
special physical theory can be used in other domains of science.
For example, we are not surprised that differential calculus which
was developed to serve for classical Newtonian mechanics was later
used in field theory, QM, biology, economics. Nobody protests
against applying the classical probability calculus (the
Kolmogorov measure-theoretic model) to modelling of financial
processes and so on. In the same way we import the mathematical
formalism of quantum mechanics to cognitive science and
psychology, even without trying to perform a reduction of mental
processes to quantum physical processes.

\medskip

To escape misunderstanding, we shall reserve notations classical and
quantum for physics. And in applications outside physics we shall
use notations {\it classical-like} (CL) and {\it quantum-like} (QL).

\medskip

In this article we shall argue that the functioning of the human
brain, in particular that responsible for and engaging with
consciousness, may be QL, without, as against the arguments
advanced by recent quantum theoretical approaches to
consciousness, necessary being physically {\it quantum.}
Specifically the QL character  of this functioning means that it
may be described by the same mathematical formalism as that used
in QM (say, in von Neumann's version of the formalism), even
though the physical dynamics underlying this functioning may not
be the same as that of the physical processes considered in QM (as
physical theory). In other words, the article offers  a new {\it
mathematical model} of the dynamics of the brain, the model based
on the same mathematics  as that used in QM, without viewing this
dynamics itself as physically quantum. In opposite to views of
quantum (physical) reductionists of brain's functioning -- e.g.
Hameroff (1994, 1998) and Penrose (1989, 1994, 2005), we are not
interested in brain's physical structure on the micro level. We do
not expect to extract cognitive features of the brain from quantum
physical processes.

\subsection{On  quantum reduction of cognition}

By using QL-models we might escape some
problems arising in the quantum reductionist approach, e.g., the
presence of the huge gap between the quantum (physical) and
neurophysiological scales.

Roughly speaking at the neuronal level the brain is too hot and
too noisy to be able to operate as a physical quantum system. Of
course, one could try to save the quantum reductionist model by
rejecting the commonly accepted neuronal model of the functioning
of the brain. One might speculate that ``real mental life'' is
going on the micro level and that the macro neuronal activity
plays a subsidiary mental role.

Moreover, we could not reject even the hypothetical
possibility that ``real mental life'' is going  on the scales of space  and time which are even
finer than those which are coupled to QM (of e.g. electrons, protons or neutrons):
Penrose speculated that consciousness is created on the scales of {\it quantum gravity.}

However, such a revolutionary approach would lead to a number of
problems. First it is not easy to accept that neurons are not at
all basic units of information processing in the brain. Both
theoretical and experimental neurophysiology give  a huge number
of evidences in favor of neural processing of cognitive
information. Then even if neural activity plays a subsidiary role
in cognitive processes comparing with basic quantum activity in
the crowed, the brain could not completely ignore the former.
Therefore we  should be able to explain how ``quantum cognition''
could be lifted from micro world to the neural level. At the
moment such ``lifting models'' do not exist.

\subsection{Can neuronal and quantum-like models be combined?}

As was pointed out the QL approach provides just a special representation of information which is described
by the mathematical apparatus of QM -- the calculus of averages which are calculated
not by using classical integration (summation in the case of discrete random variables), but by taking the
operator trace in Hilbert space.

Therefore information underlying the QL representation
can be produced by micro as well as macro systems. In contrast to ``really quantum''
reductionist theories of cognition,
a QL model can be  based on the QL representation of (purely classical) information
produced by neurons, see Khrennikov (2002b, 2004a) and Behera et al. (2005).

One can see huge difference between our QL approach and the ``really quantum''
approach. By the latter a neuron could not process quantum information, because it (as a hot and noisy
macroscopic system) could not be being in superposition of two macroscopically distinguishable states:
firing and nonfiring.\footnote{Penrose (1994):
``It is hard to see how one could usefully consider a quantum superposition consisting
of one neuron {\it firing,} and simultaneously {\it nonfiring.}''}
We do not need to consider such ``neuronal Schr\"odinger
cats.'' In a coming QL model of cognition internal states of neurons are  ``mental hidden variables''
for the QL representation.

\subsection{Conflict with the Copenhagen interpretation of quantum mechanics}

Of course, any reader who is little bit aware about
80 years of debates on the problem of so called {\it completeness of QM}
immediately understands  that we would be in trouble: the majority of  the modern quantum community believes
that QM is complete. It seems that hidden variables and in particular
{\it mental hidden variables}
could not  be introduced (even in principle).

The discussion on completeness of QM was initiated by the famous
paper of Einstein, Podolsky and Rosen (1935). In fact, Albert
Einstein was strongly against the so called {\it Copengahen
interpretation of QM.}  Such a rejection of  one special
interpretation of QM is often misinterpreted as rejection of  QM
by itself. Of course, Einstein was not against QM. However, he
considered QM as only an {\it approximative description} of
physical processes in the micro world. In particular, he was sure
that one will create (sooner or later) a more fundamental theory
which would explain so called {\it ``quantum randomness''} in
terms of classical stochastic processes.

By the Copenhagen interpretation quantum randomness
differs crucially from classical randomness. The latter is reducible to statistical variation of features
of elements of a huge ensemble of systems under consideration. The former  is fundamentally irreducible.
It is ``individual randomness'' of e.g. electron of photon.

Einstein was sure that quantum randomness can be reduced to
classical one. In particular, the wave function should be
associated not with a single quantum system, but with an ensemble
of equally prepared quantum systems. This is so called Einsteinian
or {\it ensemble interpretation} of QM. Thus for Einstein it was
totally meaningless to speak about ``the wave function of
electron'' or about ``collapse of such a wave function''. We
remind that in the late 20s and early 30s Schr\"odinger was very
sympathetic to the ensemble interpretation, see his correspondence
with Einstein in Schr\"odinger (1982).\footnote{It may be not so
well known that even the example with {\it Schr\"odin´ger's cat}
was created under the influence of Einstein to show absurdness of
the Copenhagen interpretation. Originally Einstein considered
bomb.} However, finally Scr\"odinger was about to reject not only
the Copenhagen, but also the ensemble interpretation, since {\it
Einstein was not able to explain why probabilistic formalisms of
classical and quantum physics differ so much}: on the one hand,
the integral calculus and, on the other hand, the operator
algebra. This important problem of {\it coupling of the quantum
probability calculus with the classical probability calculus was
solved} in Khrennikov (2005a,b).

\subsection{Bell's inequality}

In spite of many years of debates, there is still
no definite answer to the question which was asked in  Einstein et al (1935).
Although arguments based on Bell's inequality,
see e.g. Bell (1964, 1987) and Clauser et al (1982), and subsequent experimental studies, see  Aspect et al (1982)
and Weihs et al (1998), Weihs (2007),
induced rather common opinion that if QM were incomplete
(i.e., a deeper ``hidden variable'' description is possible), then any subquantum classical
statistical model is nonlocal. Thus one could  choose between ``death of reality'' (impossibility
to introduce hidden variables) and nonlocality.
It is interesting that not so many quantum physicist are ready to accept a classical  nonlocal
model of e.g. Bohmian type as a subquantum model. Although the ``official conclusion'' from Bell's
arguments is incompatibility of QM and local realism, in reality it is completeness
of QM (i.e., impossibility of a deeper ``subquantum'' description of micro phenomena)
that is questioned.\footnote{We remark that this conclusion totally contradicts to the initial plan of J. Bell to show
that QM is incomplete, but subquantum classical reality is nonlocal. One of my colleagues
and a former student of David Bohm pointed out at many occasions to the real abuse of original
Bell's arguments in modern quantum physics.}

At the first sight such a problem as ``death of reality'' might be
interesting only for philosophers. However, last 10 years quantum
information was intensively developed. And nowadays purely
philosophic problems of quantum foundations play  an important
role in nanotechnological projects for billions of dollars. Take
quantum cryptography as an example. Definitely completeness and
not contradiction between local realism and QM plays the crucial
role in justification of projects in quantum cryptography.
Postulated incredibly security of quantum cryptography is based on
the impossibility for Eva to create a classical probabilistic
model with hidden variables underlying quantum communication
between Alice and Bob. If such a model were possible, Eva would
enjoy reading of quantum communications between Alice and Bob. As
was pointed out, people ignore the possibility of classical, but
nonlocal attacks on quantum cryptographic schemes.

\subsection{Can a mathematical theorem be used as a crucial physical argument?}

In spite of the evident fact that the majority believes in Bell's arguments, a number of experts
actively criticise these arguments, see e.g. Pearle (1970),
Accardi and Fedullo (1982), Pitowsky (1982), Fine (1982), De Baere (1984), Gisin, N. and Gisin, B. (1999),
Klyshko (1993a,b, 1995, 1996a,b, 1998a,b),
Larsson (2000), Hess and Philipp (2001, 2002, 2006),
Volovich (2001), Khrennikov (1999b), Khrennikov and Volovich (2005),
Adenier and Khrennikov (2006), Accardi and Khrennikov (2007).
 We are not able to discuss those anti-Bell arguments in this
paper. We just point out that Bell's inequality as any mathematical theorem is based on a number of mathematical
assumptions. Some assumptions can be questioned either from mathematical or physical viewpoints.\footnote{For example,
Bell assumed that ranges of values of classical ``subquantum variables'' coincide with ranges of values
of corresponding quantum observables (for example, the classical spin variable
should take the same values as the quantum spin observable, i.e., $\pm 1).$ If we proceed without this assumption (so we consider a more complicated
picture of correspondence between classical and quantum models) then we can reproduce quantum correlations as classical
ensemble  correlations, see
Accardi and Khrennikov (2007).}

Personally I am very surprised that serious people hope that it might be possible to prove something about physical reality
on the basis of a mathematical theorem.\footnote{I recall that Einstein  completely ignored the first
known ``NO-GO'' theorem which was proved by von Neumann in early 30s. It is especially curious, since
von Neumann's book, see e.g. von Neumann (1955), was being in Einstein's office during many years.}

Can one say that non-Euclidean geometries do not exist in nature, because
Pythagorean theorem is mathematically rigorous? Of course, one could also check that geometry of space is
Euclidean  by measurements of angles in triangles. Suppose that
measurements show that geometry is Euclidean. But for what scale? Can one exclude that that geometry might be
non-Euclidean at a finer scale?

Finally, suppose that measurement devices work so badly that about 80$\%$
of triangles are simply destroyed. One should base her conclusions only on a sub-sample containing
20$\%$ of triangles. Moreover, one could not even exclude that sampling is unfair, i.e., that the measurement
devices just select Euclidean triangles and destroy non-Euclidean, cf. QM --
violation of Bell's inequality: Pearle (1970),
Gisin, N. and Gisin, B. (1999), Larsson (2000), Adenier and Khrennikov (2006).

\subsection{Quantum-like processing of incomplete information}

We decide not to wait for the end of the exciting  Einstein-Bohr debate and the complete clarification
of the theoretical and experimental situation for the EPR experiment and Bell's inequality.
We proceed  already now by creating a QL model in that the mathematical structure of QM is
reproduced. Such a model is based on an incomplete representation of
classical information.

\medskip

{\it What are advantages of
the incomplete-information interpretation of the mathematical formalism of QM? }

\medskip

In such an approach the essence of the quantum mathematical
formalism is not the description of a special class of physical
systems, so called quantum systems, having rather exotic and even
mystical properties, but the possibility to operate with {\it
incomplete information} about contexts.\footnote{In principle,
context may be represented by an ensemble of systems, but may be
not.}  Thus one can apply the formalism of QM in any situation
which can be characterized by the incomplete description of
contexts (physical, biological, mental). This (mathematical)
formalism could be used in any domain of science, cf. Aerts, D.
and Aerts, S. (1995, 2007), Khrennikov (1999a, 2000, 2002b, 2003b,
2004a, 2006a), Choustova (2001, 2004), Haven (2006), Busemeyer
 et al (2006, 2007 a,b), Franco (2007): in cognitive and social sciences, economy,
information theory. Contexts under consideration need not be
exotic. However, the complete description of them should be not
available or ignored (by some reasons).

We shall use the incomplete-information interpretation of the formalism of QM.
By our interpretation it is a special mathematical
formalism working in the absence of complete information about context.
By using this formalism a cognitive system permanently
ignores huge amount of information. However, such an information processing does not induce
chaos. It is extremely {\it consistent.} Thus the QL information cut off is done in a clever
way. This is the main advantage of the QL processing of information.

Suppose that a biological organism is not able to collect or/and to process the complete set of information
about some  context: either  because of some restrictions for observations
or because it is in hurry -- to decide immediately or to die!
Such an organism  can create a model of phenomena which is based on permanent and consistent
ignorance of a part of information. Evolution can take a large number of generations. Consistency of ignorance
plays the crucial role. Organisms in a population should elaborate the same system of rules
for ignorance of information, otherwise they would not be able to communicate.
By our interpretation the QL formalism provides the consistent rules for such a modelling of reality.
We assume that higher level biological organisms and especially human beings developed in the process of evolution
the ability of QL processing of information. Thus QL processing is incorporated in our brains.

By our approach cognitive evolution of
biological systems can be considered as evolution from purely
classical cognition (neural processing of complete data) to QL cognition.
From this viewpoint human beings are essentially more quantum than
e.g. dogs and dogs than crocodiles. Essential number of algorithms
which are performed by the human brain are QL (of course,
classical algorithms and networks also play essential roles). One
might speculate that crocodiles operate merely with classical
algorithms. Reading of e. g.  Fabre (2006) gives the strong
impression that insects' activity has no QL-elements.

\subsection{Time scales of the functioning of brain and QL model}

By our model, see section 4.2,  the QL brain works on two time scales:
subcognitive (pre-QL) and cognitive (QL). At the first (fine) time scale
information process is described by classical stochastic dynamics which is transformed
into QL stochastic dynamics at the second (rough) time scale.

To couple our model to physiology, behavioral science, and
psychology, we consider a number of known fundamental time scales in
the brain. Although the elaboration of those scales was based on
advanced experimental research, there are still many controversial
approaches and results. The temporal structure of the brain
functioning is very complex. As {\it the physiological and
psychological experimental basis} of our QL-model we chosen results
of investigations on one special quantal temporal model of mental
processes in the brain, namely, {\it Taxonomic Quantum Model} --TQM,
see Geissler et al (1978), Geissler and Puffe (1982), Geissler
(1983, 85, 87,92), Geissler and Kompass (1999, 2001), Geissler,
Schebera, and Kompass (1999). The TQM is closely related with
various experimental studies on the temporal structure of mental
processes, see also Klix and van der Meer (1978), Kristofferson
(1972, 80, 90), Bredenkamp (1993), Teghtsoonian (1971). We also
couple our QL-model with well known experimental studies, see, e.g.,
Brazier (1970), which demonstrated that there are well established
time scales corresponding to the alpha, beta, gamma, delta, and
theta waves; especially important for us are results of Aftanas and
Golosheykin (2005), Buzsaki (2005).

The presence of fine scale structure of firing patterns which was
found in Luczak et al (2007) in experiments which demonstrated
self-activation of neuronal patterns in the brain is extremely
supporting for our QL-model.\footnote{"Even in the absence of
sensory stimulation, cortex shows complex spontaneous activity
patterns, often consisting of alternating "DOWN" states of
generalized neural silence and "UP" states of massive, persistent
network activity. To investigate how this spontaneous activity
propagates through neuronal assemblies in vivo, we recorded
simultaneously from populations of 50-200 cells in neocortical
layer V of anesthetized and awake rats. Each neuron displayed a
virtually unique spike pattern during UP states, with diversity
seen among both putative pyramidal cells and interneurons,
reflecting a complex but stereotypically organized sequential
spread of activation through local cortical networks. The
timescale of this spread was ~100ms, with spike timing precision
decaying as UP states progressed," see Luczak et al (2007).} Of
course, not yet everything is clear in neurophysiological
experimental research, see Luczak et al (2007): "The way
spontaneous activity propagates through cortical populations is
currently unclear: while in vivo optical imaging results suggest a
random and unstructured process Kerr et al (2005), in vitro models
suggest a more complex picture involving local sequential
organization and/or travelling waves, Cossart et al (2003), Mao
(2001), Ikegaya (2004),  Sanchez-Vives and McCormick (2000), Shu,
Hasenstaub, and McCormick(2003), MacLean (2005)."

In any event our QL-model for brain functioning {\it operates on
time scales which are used in neurophysiology, psychology and
behavioral science. This provides an interesting opportunity to
connect the mathematical formalism of QM with theoretical and
experimental research in mentioned domains of biology.} We hope
that our approach could attract the attention of
neurophysiologists, psychologists and people working in behavioral
science to quantum modelling of the brain functioning. On the
other hand, our QL-model might stimulate theoretical and
experimental research on temporal structures of the brain
functioning.

\section{QL averages as the basis of advanced cognition}

\subsection{Classical neural averages as cognitive images}

We consider the following model of the functioning of the brain.
Brain's state $x$ is combined of states of individual neurons.
 Any mental function $f$ is realized as a function of the state: $f(x).$  We consider
$x$ as classical random vector (so mathematician would write $x=x(\omega),$ where $\omega$ is a random parameter)
describing fluctuations of neural activity.\footnote{In a more advanced model
the brain is split in a number of networks and each network determines its own state-variable. Different
mental functions are associated with different networks.} By our model the brain does not ``feel'' fluctuations
of the mental function $f(x)$ induced by fluctuations of the neural state $x.$ Brain's images are purely
statistical. The brain ``feels'' the average $\langle f \rangle$ of the mental function. Distributions
of neural activity produce probabilistic cognitive images.

This is a purely classical {\it probabilistic model of cognition,} see Khrennikov (2004a,b) and  for its neural realization.
Of course, to elaborate it one should
discover ``statistical neural code'': to make correspondence between averages of a mental function and cognitive
images corresponding to this functions. One could not exclude the possibility that different mental
functions may have different coding systems. This is a problem of the greatest complexity, cf. problem
of neural code.

However, we do not claim that an advanced brain works on the basis of such a classical statistical cognition.
The main problem is the huge dimension of the neural-state vector (at least for advanced mental functions).
To find the statistical average   $\langle f \rangle,$ the brain should perform integration over the space
having the dimension of a few billions. It is clear that it is totally impossible to proceed in this way.

\subsection{Production of QL averages-images}

If cognitive systems really use statistical encoding of images they should find some approximative
representation of averages. We shall see that one of  simplest  approximations
of the classical average coincides with the quantum average. The latter is given by the operator-trace
formula, see von Neumann (1955).

By our approach the QL approximation  was developed by cognitive systems to speed
the process of calculation of averages with respect to fluctuations of neural activity.

\subsection{Quantum-like processing of information as a characteristic feature of advanced brains}

If the brain does not have so many neurons and neural networks it can proceed purely classically.
It can use the probabilistic representation of information which is given by
classical averages: normalized sums of values of a mental function $f.$ Therefore QL cognitive processing of information
is profitable only for ``advanced brains.'' As was pointed out in introduction, primitive cognitive
systems need not process information by using QL algorithms, because they need not even represent
information in the QL way. It is clear that QL evolution of biological organisms should produce
a {\it smooth scale} of the QL ability for information processing. In the processes of evolution
brains became more complicated on the neural level. To speed information processing,
brains should look for some approximative representation. In this way they develop the QL ability. However,
they still preserve a number of mental functions which are based on classical probabilistic algorithms.
In particular, in the human brain classical mental functions peacefully coexist with QL mental functions
(which were created at the latter stage of cognitive evolution).\footnote{One should not think that
all classically realized mental functions are  ``old'' and only QL realized mental functions
are ``new'' (relatively). If a mental function $f$ need not so much neural resources (its state
space is rather small), then there are no reason to realize $f$ by using the QL representation.
Thus even ``new'' but simple mental function can have the classical realization.}

At some moment in the process of cognitive evolution  the crucial step toward QL computing was done.
Thus our QL cognitive model induces an interesting problem of cognitive evolution:

\medskip

{\it ``What was the first biological organism with QL
information processing?''}

\medskip

Since our QL model has not yet been  sufficiently elaborated
(neither theoretically nor experimentally), it is too early to try
to give a definite answer to this question. We shall  speculate
little bit. It seems that insects are purely classical, see again
e.g. Fabre (2006), as well fishes. What is about mammals,   e.g.
dolphin? The brain of dolphin is very advanced as a neural system.
However, it may still be, although very advanced, purely
classically designed. My conjecture is that the crucial step
toward QL processing of cognitive information was done by human
beings. And this ``QL discovery'' is the main reason for rather
special cognitive evolution of human beings.

Shortly, evolution of the brain can be characterized by three
stages:

\medskip

1) deterministic neural network processing -- images are states of single neurons;

2) probabilistic processing -- not individual states of neurons, but their probability distributions
determine cognitive images;

3) discovery and development of the QL representation.

\medskip

At the first stage, the brain was fine with a relatively small number of neurons. At the second stage
the brain was interested to expand as much as possible (more neurons, more neural networks) --  by a very simple
reason: a larger number of probability distributions can be realized on  a larger state space. Hence
such a classically more complicated brain is able to create more cognitive images.
However, richness of the cognitive representation was approached by the cost of  calculation of averages
over spaces of huge dimension. Discovery of the QL representation solved this computational problem.
Now cognitive images are processed essentially faster than classical cognitive images.

The QL brain does not demand so much computational resources as it was at the pre-QL stage of development.
Moreover, QL images are less detailed than the classical ones. We recall that QL representation
is an approximative representation. Thus some ``details of the classical representation''
disappear in the process of transition to the QL representation. We shall see in section 3 that different classical
probability distributions of neural activity can be mapped into the same  QL probability distribution.
In contrast to the classical (probabilistic) brain,  the QL brain
need not be extremely precise in determining of probabilistic distributions for neural activity.
Such unsharpness of cognitive images  also implied liberation of computational resources.
Moreover, the QL advanced brain can proceed with a smaller neural state space.
It can reduce the number of neurons comparing with
the second stage of brain's evolution. The advanced QL brain can be smaller than the advanced
classical (probabilistic) brain.

The homo-brains really evolved in this way: the brain of homo
 was 10$\%$ greater than the modern human average. By
our model homo neanderthalensis was the highest stage of
extensional development based on the classical probabilistic
model. Of course, we can not exclude that such a brain already had
some elements of the QL representation. However, homo sapiens
developed quicker in the QL way. Therefore homo sapiens was able
to proceed better with a smaller brain. Our QL model might explain
the evolutionary superiority of homo sapiens comparing with homo
neanderthalensis. Otherwise it is not easy to explain extinction
of the latter who was physically stronger and who had more
extended neural state space.

\section{The basic mathematical formula for the QL approximation of classical averages}

This is the only section in our paper in that we really can not escape some mathematics.

\subsection{Covariance matrix}

We consider the average of the random vector $x.$ If this vector has coordinates
$x=(x_1,...,x_N)$ then its average is the vector with coordinates
$m_x=(m_{x_1},...,m_{x_N}),$ where $m_{x_i}= E x_i$ is the average of
the coordinate $x_i.$

We recall that the covariance matrix of a random vector is defined as the matrix
which elements are covariances between coordinates of this random vector.
$$
\rho=(\rho_{ij}), \;\rho_{ij}= E (x_i-m_{x_i}) (x_j - m_{x_j}).
$$
We recall that any covariance matrix is symmetric: $\rho_{ij}=\rho_{ji}$ and it is positively defined:
$$
(\rho x, x) = \sum_{ij} \rho_{ij} x_i x_j \geq 0.
$$
We also remark that its trace
$$
\rm{Tr} \rho = \rho_{11} + ...+ \rho_{NN}
$$
is equal to {\it dispersion} of the random vector $x: Dx= \sigma_x^2,$
where $\sigma_x$ is so called standard quadratic deviation of $x.$

\subsection{Taylor formula}

Let $f(x)$ be a smooth function. We recall that (in the multidimensional case)
its derivative at some fixed point $a=(a_1,...,a_N)$ is given by its gradient consisting of the derivatives
with respect to coordinates (partial derivatives)
$$
f^\prime(a)=\Big(\frac{\partial f}{\partial x_1}(a), ..., \frac{\partial f}{\partial x_N}(a)\Big).
$$
The right geometric interpretation of the gradient is that it is a so called covector.
There is well defined pairing with vectors:
$$
(f^\prime(a),x) = \frac{\partial f}{\partial x_1}(a) \; x_1 +...+ \frac{\partial f}{\partial x_N}(a) \; x_N .
$$
The second derivative is given by the matrix of partial
derivatives of the second order which is called Hessian:
$$
f^{\prime \prime}(a)=\Big(\frac{\partial^2 f}{\partial x_i \partial x_j}(a) \Big).
$$
We point out that this matrix is symmetric.

By using the Taylor formula of the second order in the point $a$ we get:
\begin{equation}
\label{M2}  f(x) \approx f(a) + (f^\prime(a), x -
a) + \frac{1}{2} (f^{\prime \prime} (a) (x- a), (x- a)).
\end{equation}
If we now take the average of both parts of this approximative equality by choosing
$a=m_x$ we obtain:
\begin{equation}
\label{M2X}  \langle f \rangle \equiv  Ef(x)
\approx f(m_x)  + \frac{1}{2} E (f^{\prime \prime}(m_x) (x- m_x), (x- m_x)).
\end{equation}
The term with the first derivative disappeared, because $E(x-m_x)=0$ (the operation of averaging is
linear). Thus only the second derivative is important for such a probabilistic approximation.
By using little bit of linear algebra we can represent the last term as the matrix-trace:
$
\rm{Tr}\; \rho \; A,
$
where the matrix $A=\frac{1}{2} f^{\prime \prime} (m_x)$ and $\rho$ is the covariance matrix
of the random vector $x.$

\subsection{Normalization of random variable}

In probability theory it is convenient to operate with normalized random variables. We set
$$
y= \frac{x-m_x}{\sigma_x}
$$
Then its average is equal 0 and dispersion is equal 1.
In particular, for its covariance matrix the trace is equal 1.

We now consider this normalization of the random state-vector in the approximation formula (\ref{M2X}).
By assuming that a mental function is chosen in such a way that $f(0)=0$ (if it was not so from the very beginning,
we just set $f(x) \to f(x) - f(0))$ we get the following simple approximation
formula:
\begin{equation}
\label{M2X1}  \langle f \rangle \approx \rm{Tr}\; \rho \; A,
\end{equation}
where $A= \frac{1}{2} f^{\prime \prime} (0).$

\subsection{Quantum averaging}

The quantum averaging procedure is based on linear algebra.\footnote{In fact, from the mathematical viewpoint
it is even simpler than the classical averaging procedure. The latter is heavily based on theory
of Lebesgue integration which is not so simple. The common opinion that quantum probability is
more complicated than classical probability is not a consequence of the use of more advanced
mathematics, but of counter-intuitiveness of quantum probability.}  To simplify considerations,
we discuss QM with finite-dimensional state space. Such a quantum model is basic
for e.g. quantum information: one qubit  has the two dimensional state space, $N$ qubits have
the $2^N$ dimensional state space.

Quantum observables are given by symmetric matrices. Quantum states are given by von Neumann
density matrices\footnote{We remind that Landau introduced them even earlier than von Neumann.}:
symmetric, positively defined and having the unit trace.

The quantum average of an observable $A$ with respect to a state $\rho$ is given by the
von Neumann trace formula:
\begin{equation}
\label{M2X2}  \langle A  \rangle = \rm{Tr}\; \rho \; A,
\end{equation}

\subsection{Wave function or density matrix?}

Historically quantum states are associated with wave functions.
These are so called pure states. So called mixed states are given
by von Neumann density matrices. First we remark that any pure
state $\psi$ also can be represented by a density matrix, namely,
by the orthogonal projector $\rho_\psi$ onto the vector $\psi.$
Thus formally one can proceed by using only density matrices.
However, the use of pure states has a deep interpretational basis.
By the Copenhagen interpretation $\psi$ describes the state of
{\it an individual quantum system.} For example, one speaks about
the wave function of electron. On the other hand, by the ensemble
interpretation there is no meaning to distinguish ``pure states''
and ``mixed states.'' The $\psi$-function (in fact, the
corresponding projector) as well as an arbitrary density matrix
$\rho$ describes an ensemble of specially prepared systems. Thus
$\rho_\psi$ describes a mixture as well. Of course, this is a
mixture with respect to hidden variables. By the Copenhagen
interpretation such variables do not exist at all.\footnote{At the
last V\"axj\"o conference, ``Quantum Theory: Reconsideration of
Foundations -- 4'' Arcady Plotnitsky pointed out that the
Copenhagen interpretation should not be directly identified with
Bohr's interpretation. In fact, Bohr discussed only measurements
and not the possibility to introduce hypothetical hidden
variables. For him QM was complete from the experimental
viewpoint. It was impossible to perform ``better measurements''
than those described by QM. The more orthodox interpretation that
even in principle   it is impossible to assume existence of a
deeper level of the description of nature was elaborated by Fock
(``the Leningrad version of the Copenhagen interpretation''). In
cognitive modelling such a difference in the interpretations plays
the crucial role. If mental hidden variables can exist, then they
even can be measured. Bohr might be not strongly against this
possibility, because mental hidden variables are macroscopic.}

Since we follow the ensemble interpretation, we would not speak about  ``the wave  function of the brain''
(hence, for us ``collapse of brain's wave function'' is totally meaningless phrase). Instead of this,
we shall consider QL states of brain given by density matrices. As we shall see, such mental states
arise as special representations of classical neural statistical states.

This is just the question of simplicity. For the brain it is easier to operate with QL mental states
(density matrices) than with classical probability distributions of neural activity.

\subsection{Quantum-like approximation of classical averages}

If we compare the formulas (\ref{M2X1}) and (\ref{M2X2}) we see
that quantum averaging coincides with the approximation of classical averages which was obtained
with the aid of the Taylor formula. We suppose that the brain uses the QL representation in this way.
It proceeds with quantum averages, because it is simpler to calculate them. But it does not
completely lose the ``neural ground'', because it uses approximations of real neural averages.

Thus in the QL representation the probability distribution of neural activity
is represented by its covariance operator (we recall that we operate with normalized
random state vector). The mental function $f$ is represented by its second derivative (Hessian).

The QL representation is extremely incomplete. A huge class of probability distributions
is mapped onto the same ``covariance-density matrix'' $\rho.$
In our model abstract cognitive images are created  due to the QL representation by identification
of a class of classical probability distributions of neural activity.

\section{Precision of quantum-like approximation of classical averages}

\subsection{Mental and neural realities}

One can ask: ``What is the precision of the QL approximation, see
(\ref{M2X1}), of classical averages?'' This is not simply a
mathematical  problem of the precision of some approximation
algorithm. This problem plays the fundamental role in our QL
cognitive modelling.

If approximation is very good, then the QL approximation (given by the von Neumann
trace formula) does not differ so much from the ``real average'' corresponding to  neural activity.
In such a case the use of the QL representation just make faster operation
with averages-images (because it is easier to calculate trace than the Lebesgue
integral). However, the QL representation would not deform essentially the
classical (neural) picture of reality.

If the Taylor approximation is rather rough, then the QL approximation
differs essentially from the ``real average.''
In such a case the use of the QL representation not only make faster operation
with averages-images, but it also deforms essentially the
classical (neural) picture of reality. The brain created new QL ``mental reality''
which does not have direct relation to neural reality (although ``mental reality'' is
still not independent from neural reality).

\subsection{Time scales and the precision of the quantum-like approximation}

In Khrennikov (2006d) it was shown that one can couple the precision of the
QL approximation with time scales of classical neural processing (subcognitive)
and QL processing (cognitive). We would not like to use too much mathematics in this paper
which is oriented to a multi-disciplinary auditorium. Therefore we just recall the main idea
of Khrennikov (2006d) considerations.

We start with the basic stochastic process $x(s, \omega)$ (where $s$ is the
subcognitive time and $\omega$ is the chance variable)  generated by neural activity. This process proceeds
on its own time scale. We call it subcognitive (or sub-QL) time scale:
$s_{\rm{pcogn}}.$ Classical neural averages are created as the result of fluctuations
at this time scale.

Cognition is ``not interested'' in individual fluctuations which take place at this time scale.
It is not interested in corresponding fluctuations of a mental function: $f=f(x(s,\omega)).$
It is interested only in the average of the mental function and, moreover, QL cognition uses
only the QL approximation of the average. Therefore cognition has its own time scale, {\it
cognitive time scale} (or QL-scale): $t_{\rm{cogn}}.$ The interval $s_{\rm{pcogn}}$ is
small with respect to the  cognitive time scale (it is an ``instant of time'' in the cognitive scale).
Thus cognitive images-averages are processed on the cognitive time scale -- $t_{\rm{cogn}}$ and
neural images on the subcognitive time scale $s_{\rm{pcogn}}.$

Under some assumptions on the underlying  neural stochastic process (e.g. for the process of Brownian
motion or more general Gaussian processes) we can prove that the QL approximation (the trace)
deviates from the real average (integral) by the term of the order $\kappa,$ where the latter parameter
determines the relative size of the subcognitive and cognitive scales:
\begin{equation}
\label{GH}
\kappa= \frac{s_{\rm{pcogn}}}{t_{\rm{cogn}}}.
\end{equation}
It provides a numerical measure of deviation of the QL (fuzzy, unsharp)
representation of information from  the CL (complete, sharp) representation.

Under the assumption that the subcognitive time scale $s_{\rm{pcogn}}$
is fixed, we find that for small periods of fluctuations
$t_{\rm{cogn}}$ the parameter $\kappa$ is very large. Thus {\it higher
frequencies (at the cognitive time scale) induce larger deviations
from the (complete) CL-processing of information.}
Huge amounts of information which are processed at the subcognitive
time scale are neglected, but not arbitrary (randomly). There is the
QL-consistency in the information processing.

Consequently, for low frequencies (oscillations with long periods) this coefficient is
small. Therefore the QL-processing does not imply large deviations
from the CL-computational regime.

One of the fundamental consequences of our QL model for neurophysiology and cognitive science is that

{\it Conscious (QL) processing of mental information is associated with relatively high frequencies.}\footnote{In this
context ``relatively'' is with respect to a subcognitive scale. Thus the crucial role is played not by
absolute values of frequencies, but by the gap between sub-QL and QL time scales.}

The crucial problem is to find those biological time scales which
induce the QL-representation of information.  We split the problem into the two parts:

\medskip

1) to find the subcognitive time scale;

2) to find the cognitive time scale.

And in reality the problem is even more complicated.
Our fundamental assumption is that {\it there exist various pairs of
scales inducing various QL-representations of information.}
We could not hope to find once and for ever defined pair of time scales.
Various mental functions can be based on various pairs of subcognitive and
cognitive time scales.

\subsection{Cognitive time scale: neurophysiological and cognitive data}

It seems that (as in physics, see section 6) the first problem is more complicated.
First we consider the second one. We start  the discussion on the
choice of the cognitive time scale by considering experimental
evidences,  that a moment in {\it psychological time} correlates
with $\approx 100$ msec of physical time for neural activity. In such
a model the basic assumption is that the physical time required for
the transmission of information over synapses is somehow neglected
in the psychological time. The time ($\approx 100$ msec) required for
the transmission of information from retina to the inferiotemporal
cortex (IT) through the primary visual cortex (V1) is mapped to a
moment of psychological time. It might be that by using
$t_{\rm{cogn}}=100$ msec,we shall get the right cognitive time
scale.

However, the situation is not so simple even for the second problem.
There are experimental evidences that the temporal structure of
neural functioning is not homogeneous. The time required for
completion of color information in V4 ($\approx 60$ msec) is shorter
that the time for the completion of shape analysis in IT ($\approx
100$ msec). In particular it is predicted that there will be under
certain conditions a rivalry between color  and form perception.
This rivalry in time is one of manifestations of complex level
temporal structure of brain.

Many cognitive architecture models, e.g., John Anderson's ACT-R model, Anderson (2007),
assume that each computation
step (production rule firing) takes $\approx 50$ msec. This accounts for data well.
It might be that by using
$$
t_{\rm{cogn}} \approx 50-100 \; \; \rm{msec}
$$
we shall get the right scale of the QL-coding.

\section{Quantum-like and classical-like regimes of brain's functioning}

 As was already mentioned, if the scaling parameter $\kappa$ is very
small the brain does not lose too much  classical information.
This is practically the CL-computation. But if $\kappa$ is rather
large, then the brain works in a nonclassical regime. One may say
(cf. Birkhoff and von Neumann (1936)) that in such a regime the
brain uses {\it nonclassical logic.} However, in our approach the
brain ``produces QL logic'' on the basis of purely classical
(Boolean) logic of neural processing.

In such a QL process huge amounts of information are permanently
neglected. But this does not generate a kind of chaos. Information
is neglected in the consistent QL way.

As was pointed out a few times, such a QL-processing of information
save a lot of computational resources. It might be an important
factor of the natural selection of biological organisms.

\section{Quantum and subquantum physical models}

We now come back to physics (which was the starting point of our
QL modelling). Here we can proceed in the same way and consider
two time scales, Khrennikov (2006d). One scale, we call it {\it
subquantum,} is a fine time scale, another, we call it {\it
quantum,} is  a coarser time scale. Oscillations at the subquantum
time scale are averaged. However, the conventional QM is not about
such subquantum classical averages. It is a calculus of
approximative averages given by (\ref{M2X1}). Such an
approximation works well at the quantum time scale. The latter
time scale is considered as {\it an observational time scale.} The
scale of observations in modern laboratories. Therefore it is
natural to choose it as {\it the atom time scale:}
$$
t_{\rm{q}} \approx 10^{-21} \; \rm{sec}.
$$
The problem of the choice of a subquantum scale is more complicated.
One of possible choices of  a subquantum time scale is the
{\it Planck time scale:}
$$
s_{\rm{subq}}\approx 10^{-44} \; \rm{sec}.
$$
We remark that for such a choice of time scales the scaling parameter
$$
\kappa \approx 10^{-23}
$$
is negligibly small. On the one hand, such a good approximation explains the huge predictive power
of QM. On the other hand, it makes practically impossible experimental tests
which would show the deviation of the QM-predictions from the predictions of
the subquantum model: namely, the deviation of the trace average given by the QM
formalism from the real average obtained on the basis of experimental data.

Thus if the above choice of the time scales was done properly, then our model
has only  theoretical value for physics. However, in cognitive science the gap between scales is not so huge.
Therefore we can expect visible effects of QL behavior.

\section{Variety of  time scales in brain and quantum-like cognitive
representations}

The main lesson from the experimental and theoretical investigations
on the temporal structure of processes in brain is that there are
various time scales. They  correspond to (or least they are coupled
with) various aspects of cognition. Therefore we are not able to
determine once and for ever the cognitive time scale $t_{\rm{cogn}}$
(``psychological time''). There are few such scales. We shall
discuss some evident possibilities.

\subsection{On variety of quantum and subquantum time scales in physics}

Before to go deeper in the temporal structure of mental processes,
we shall analyze in more detail the multi-scale temporal aspects of
QM. Such aspects have never been discussed.
On the one hand, it was commonly assumed that QM is
complete (this is the Copenhagen interpretation). On the other
hand, the quantum formalism is used by only one class of observers
-- human beings who discovered the mathematical formalism of QM for one special observational
(quantum) time scale.

However, we can consider a possibility that
there exits a class of observers ("super-clocks civilization") which
use a time scale $t_{\rm{q}}^\prime$ which is essentially finer than our
time scale $t_{\rm{q}}:$
$$
t_{\rm{q}}^\prime << t_{\rm{q}}.
$$
Thus such a civilization approached another time scale in its laboratories which is essentially better
than atomic time scale.

Suppose that the super-clocks civilization has also created QM -- a special
probabilistic representation of information about measurements. Of course, its time scale
should not be extremely fine comparing with the subquantum  time
scale $s_{\rm{subq}}:$
$$
s_{\rm{subq}} << t_{\rm{q}}^\prime
$$
(we assume that both civilizations -- our and super-clocks -- are
interested in processes at the same subquantum time scale). The
super-clocks civilization would discover the same mathematical
formalism of QM. But the presence of deviation from
subquantum reality would be more evident with respect to their time
scale (since $s_{\rm{subq}}$ is the same, but $t_{\rm{q}}^\prime$ is smaller
than $t_{\rm{q}},$ the coefficient $\kappa^\prime$ for the super-clocks
civilization is larger than the coefficient $\kappa$ for our
civilization).

On the one hand, the super-clocks civilization has
a better possibility to find deviations of the incomplete quantum
description from the complete classical
. However, there
might be chosen a strategy to ignore such deviations and still use
the quantum picture of the world. Even if it does not match
precisely with the complete set of information about external world,
it might be, nevertheless, convenient (by computational and
consistency reasons) to proceed with the quantum pictures of
reality.

\subsection{Temporal structure of the brain}

By our QL model a similar functioning with a few time scales
is present in the brain. How can we find those
scales?

It is well known, see, e.g., Brazier (1970), that there are well
established time scales corresponding to the alpha, beta, gamma,
delta, and theta waves. Let us consider these time scales as
different cognitive scales. There is one technical deviation from
the QL-scheme which was discussed above. We cannot determine
precisely definite cognitive times corresponding to these scales.
The scales are defined by ranges of frequencies and hence ranges of
scaling times.

For the alpha waves we choose its upper limit frequency, 12 Hz, and hence the
$t_{\rm{c}, \alpha} \approx 0.083$ sec. For the beta waves we consider (by taking upper bounds of
frequency ranges) three different time scales:
15 Hz, $t_{\rm{c}, \beta, \rm{low}} \approx 0.067$ sec. -- low beta waves,
18Hz, $t_{\rm{c}, \beta} \approx 0.056$ sec. -- beta waves,
23 Hz $t_{\rm{c}, \beta, \rm{high}} \approx 0.043$ sec. -- high beta waves. For gamma waves
we take the characteristic frequency 40 Hz and hence the time scale
$t_{\rm{c}, \gamma} \approx 0.025$ sec.

The gamma scale is the finest and hence processes represented at
this scale has the highest degree of QL-ness. On the other hand, we
know that gamma waves patterns in the brain are associated with
perception and consciousness. The beta scale is coarser than the
gamma scale and it has less degree of QL-ness in processing of
information.  We know that beta states are associated with normal
waking of consciousness.

The theta waves are even less QL than the alpha waves. They are
commonly found to originate from occipital lobe during periods of
relaxation, with eyes closed but still awake.  They  are involved
into a representation of information with  a high degree of
classicality. And these rhythms are observed during some sleep
states, and in states of quiet focus, for example, meditation,
Aftanas and Golosheykin (2005). However, there are also experimental
evidences that the theta rhythms are very strong in rodent
hippocampi and entorhinal cortex during learning and memory
retrieval. We can just speculate that learning needs using of an
essentially more detailed information representation. Thus learning
(or at least a part of it) is less QL and hence more CL. The same we
can say about memory retrieval. It also needs a more complete,
CL-representation of information. Large body of evidence, Buzsaki
(2005), indicates that theta-rhythms are used in spatial learning
and navigation. Here we present the same reasons: such tasks are
based on the CL-representation of information.

Finally, we consider delta waves. Comparing with the highest scale
-- the gamma scale, the delta time scale is extremely rough. This
induces a low degree of QL-ness. This is the state of deep
sleep.\footnote{The phenomena of sleep and dreaming are extremely
complicated. We do not plan to study them  in this paper.}

Although we still did not come to the difficult problem, namely,
determination of the subcognitive time scale, we can, nevertheless,
compare the degree of QL-ness of various time scales.

Our choice of the subcognitive time scale will be motivated by so
called {\it Taxonomic Quantum Model}, see Geissler et al (1978),
Geissler and Puffe (1982), Geissler (1983, 85, 87,92), Geissler and
Kompass (1999, 2001), Geissler, Schebera, and Kompass (1999),  for the
representation of cognitive processes in the brain (which was
developed on the basis of the huge experimental research on
time-mind relation, see also Klix and van der Meer (1978),
Kristofferson (1972, 80, 90),  Bredenkamp (1993), Teghtsoonian
(1971). In the following section we recall briefly the main features
of this model.

\section{Taxonomic quantum model}

There could be presented a portion of good criticism against
starting from EEG bands. Indeed, this band structure is one of the
few indications that directly point to behaviorally relevant
physiological properties. Physiologists suggesting the definitions
had a good intuition. However, that these definitions depend on
behavioral information is shown by enormous individual differences
in the band structures that can be defined only on a behavioral
basis. To some degree this concerns also the general band structure.
Because of individual differences, alpha is often restricted to the
"common" range which is too short to be theoretically fully
relevant. Definitions often go only from 9 to 12 Hz.  Most careful
investigators (earliest Livanov) defined the band by the range 7.5
to 13.5 Hz.

Therefore we propose to start with Taxonomic Quantum Model (TQM),
Geissler et al (1978), Geissler and Puffe (1982), Geissler (1983,
85, 87,92), Geissler and Kompass (1999, 2001), Geissler, Schebera,
and Kompass (1999). Why do we propose to use
 TQM for start of theory instead of, say,
some characteristic physiological parameters such as neuronal
refractoriness, transmission times, coupling strength etc.? In my
view, the reason is that {\it the only basis for interpreting
physiological facts of brain processes are psychophysical
(behavioral) observations, either based on motor reactions of
conscious beings or verbal reports on conscious events.} This was
the main way of thinking of von Bekesy (1936). Of course, many of
the functional statements of physiologists have the same basis. For
our purpose, this statement is absolutely essential, because a
coherent account of temporal properties of brain activity must not
only be related to behavioral observations, but it must be based on
temporal invariants extracted by a coherent theoretical account of
behavioral observations, and only these can provide the guideline to
find the proper physiological correspondences.

The best short cut to the approach is through the history of its
emergence: The first impulse towards a taxonomic turn arose in the
early 1970s from the discontent of  Geissler, see, e.g., Geissler et
al (1978), with the fact that in simple psychophysical tasks data
could indistinguishably be fitted to models resorting to widely
differing, often enough even contradicting, assumptions. In his
research in visual recognition, to circumvent this difficulty,
Geissler introduced a technique of "chronometric cross-task
comparison. " The main idea was to disambiguate models by temporal
parametrization, thereby postulating invariance of time parameters
under variation of stimulus parameters and task constraints (see
e.g. Geissler et al. (1978) and Geissler and Puffe (1982)). At that
time another research group at the same institute did something
similar by fitting latencies in standardized reasoning tasks to
predicted numbers of operations, e.g., Klix and van der Meer (1978).
The estimates from the two lines of studies yielded a surprising
picture: There seemed to exist small "bands" of operation times
centering at around 55, 110 and 220 ms, thus exhibiting
near-doubling relations. As a datum from the literature which fitted
into this regularity the asymptotic value of 36.5 ms determined by
Kristofferson (1972), see also  Kristofferson (1980, 90),  came to
mind which up to the first decimal is 1/3 of 110 ms. Taken together,
these four values suggested a system of "magic numbers". Herein a
period of 110 ms represents something like a "prototype duration"
from which the rest of periods derives by either integer division or
multiplication. From various fit procedures for step lengths,
Buffart and Geissler came up with an largest common denominator
(l.c.d.) of 9.13 ms (see Geissler, 1985) showing a standard
deviation of 0.86 ms across individuals. It turned out that the four
above-mentioned periods, although partly many times larger than this
small period, can be represented as integer multiples of it, with
nearly absolute precision: $4 \times 9.13 = 36.5; 6  \times 9.13 =
54.8; 12  \times 9.13 = 109.6; 24  \times 9.13 = 219.1.$ Of course,
this might have been some strange coincidence. Yet, later,
chronometric analysis seemed to support a modular unit of some 9 ms
(see Geissler (1985); Puffe (1990); Bredenkamp (1993). Further
investigations justify a modified assumption about quantal graining:

\medskip

{\it Regression yields the largest common denominator (l.c.d.) 4.6
ms, which is nearly exactly one half of 9.13 ms.}

\medskip

Note that, in terms of hypothetical quanta, a period of such duration represents the
next smaller candidate of a "true" elementary "time quantum" which is compatible
 with the recognition data. In the following, let us adopt provisionally the
("ideal") value of
$$
Q_0 = 4.565 \rm{ms}
$$  for this time quantum hypothesis.

The solution TQM offers to these seeming contradictions, see
Geissler (1987, 92, 85) can be considered as a generalization or at
least an analogue of the psychophysical principle of relative-range
constancy.   According to Teghtsoonian (1971), this principle
expresses itself in the fact that for all sensory continua, in terms
of output magnitudes, the ratio of the largest to the smallest
quantity is a constant of around 30. About the same value is
obtained from the so-called Subjective Weber Law.

The generalization of the principle in the realm of quantal timing
is the quantal-range constraint. To see how this analogue reads, consider
first the assumed smallest period $Q_0.$ For integer multiples
$n \times Q_0,$
consistency with the relative range constraint implies $n \leq M,$ with $M$
being a constant of the hypothetical value 30. It follows that periods of durations in
excess of $30 \times Q_0 \approx 137$ ms cannot be represented within this smallest
possible range. To account for such periods, we have to assume larger ranges
with correspondingly larger admissible smallest quantal periods to be operative.
To retain consistency with the time quantum assumption, these periods must be
integer multiples of $Q_0$ or, formally,
\begin{equation}
\label{TR}
Q_q = q \times Q_0
\end{equation}
with integer $q$ must hold. Thus, in general, the maximum extension
of any quantal of periods $T_i$ belonging to it is given by
$q \times Q_0 \leq T_i \leq M \times q \times Q_0.$
Note that the lower bound $q \times Q_0$
also defines the smallest possible distance between admissible periods within a range.
For this reason we will speak of it as the quantal resolution within a given range.
Of course, in the actual development, this abstract definition resulted from a
variety of empirical relationships suggesting a range ordering of quantal periods
with upper bounds maximally at 30 times the value of quantal resolution.

TQM does not exclude the possibility that there can be found smaller
characteristic time scales, e.g., $Q_0/30.$

\section{On the choice of subcognitive time scale}

We choose $Q_0$ as the unit of the subcognitive time:
\begin{equation} \label{PC} s_{\rm{pcogn}}=Q_0=4.6 ms
\end{equation}
This corresponds to frequencies $\approx 220$ Hz. Under such an
assumption about the subcognitive scale we can find the measure of
QL-ness for different EEG bands. For the alpha scale, we have
$$
\kappa_\alpha= \frac{Q_0}{t_{\rm{c}, \alpha}}\approx 0.055.
$$
For the beta scales, we have:
$$
\kappa_{\rm{c}, \beta, \rm{low}}= \frac{Q_0}{t_{\rm{c}, \beta, \rm{low}}}\approx 0.069;\;
\kappa_{\rm{c}, \beta}=\frac{Q_0}{t_{\rm{c}, \beta}}\approx 0.082;\;
\kappa_{\rm{c},\beta,\rm{high}}= \frac{Q_0}{t_{\rm{c},\beta,\rm{high}}}\approx 0.107.
$$
For the gamma scale we have:
$$
\kappa_\gamma= \frac{Q_0}{t_{\rm{c}, \gamma}}\approx 1.84.
$$
Thus QL-ness of processing of information increases. ``Thinking
through the alpha waves'' is more likely processing of information
by ordinary computer. Not so much information is neglected.
Therefore the information processing is not so tricky: there is no
need to manipulate with extremely incomplete information in the
consistent way. ``Thinking through the gamma waves'' is similar to
processing of information by an analogue of quantum computer --
QL-computer, see Khrennikov (2006a). Such an information processing
is very tricky: permanent informational cuts, but in the consistent
QL-way. Finally, we come to the theta and delta scales. For the
theta scale $t_{\rm{c}, \theta}= 0.125$ sec. Thus
$$
\kappa_\theta= \frac{Q_0}{t_{\rm{c},\theta}}\approx 0.037.
$$
And for the delta scale $t_{\rm{c}, \delta}= 0.5$ sec and hence:
$$
\kappa_\delta= \frac{Q_0}{t_{\rm{c},\delta}} \approx 0.009.
$$
Here the difference between the biological QL-processing of information  in the brain
and the CL-processing (as in models of artificial intelligence)  is practically
negligible.

We now compare our QL-scales of time with the "quantum scales" which
were chosen in Khrennikov (2006d):
\begin{equation}
\label{PCX} s_{\rm{pcogn}}\approx 10^{-3} \; \rm{sec}, \;  \;  \;
s_{\rm{pq}}\approx 10^{-44} \; \rm{sec}.
\end{equation}
\begin{equation}
\label{PCX1} t_{\rm{cogn}}= 30 Q_0\approx 10^{-1} \; \rm{sec}, \;  \;
\; t_{\rm{q}}\approx 10^{-21} \; \rm{sec}.
\end{equation}
Thus our model is based on macroscopic time scales, in the
opposition to really quantum reductionist models.

If we follow TQM in more detail then we should consider a
possibility that in the brain there exist a {\it hierarchy of
subcognitive times,} i.e., the above model with one fixed
subcognitive time given by (\ref{PC}) was oversimplified. From the
point of view of TQM each $Q_q$ given by (\ref{TR}) could serve as
the basis of a subcognitive time scale. We obtain a picture of
extremely complex QL-processing of information in the brain which is
based of the huge multiplicity of various subcognitive/cognitive
scales.

In this framework the notion
``subcognitive'' loses its absolute meaning. The notions
``subcognitive''/``cognitive'' become
relative with respect to a concrete psychological function
(cognitive task). Moreover, a time scale which is subcognitive
for one psychological function can be at the same time cognitive for another.

But the crucial point is that the same cognitive time scale, say $t_{\rm{cogn}},$ can
have a number of different subcognitive scales:
$$
Q_{q_1} \leq ... \leq Q_{q_m}.
$$
Each pair of scales
$$
(Q_{q_1}, t_{\rm{cogn}}),..., (Q_{q_m}, t_{\rm{cogn}})
$$
induces its own QL-representation of information. Therefore the same
$t_{\rm{cogn}}$-rhythm can be involved in the performance of a few
different psychological functions.

The final message from TQM is that the cognitive time $t_{\rm{cogn}}$
scale should be based on an integer multiplier of the time quant
$Q_0:$
\begin{equation}
\label{MP}
t_{\rm{cogn}}= N Q_0.
\end{equation}
In such a model we can totally escape coupling with directly defined
different EEG bands, alpha, beta, gamma,... We shall use only
behaviorally defined time scales. The Weber law gives us the
restriction to the value of the multiplier: $N \leq 30.$

\bigskip

{\bf References}

Accardi, L. and Fedullo,  A., 1982. Lett. Nuovo Cimento 34, 161.

Accardi, L., 1995. Il Nuovo Cimento B  110, 685.

Accardi, L. and Khrennikov, A.Yu., 2007. Chameleon effect, the range of values hypothesis and
reproducing the EPR-Bohm correlations. In:  Foundations of Probability and
Physics-4, Adenier, G., Fuchs,  C.  and Khrennikov, A. Yu.,  eds.,
American Institute of Physics,  889, Melville, NY, 21-29.

Adenier, G. and Khrennikov, A.Yu., 2006. Anomalies in EPR-Bell experiments. In:  Quantum theory:
reconsideration of foundations---3, Adenier, G., Khrennikov,  A. Yu.  and Nieuwenhuizen, Th.M., eds.,
American Institute of Physics,  810, Melville, NY,  pp. 283--293.

Aerts, D. and Aerts, S., 1995.   Applications of quantum statistics in psychological
studies of decision-proceses.  Foundations of Science,
1, 1-12.

Aerts, D. and  Aerts, S., editors, 2007. Optimal Observation: From
Quantum Mechanics to Signal Analysis (Einstein Meets Magritte: An
Interdisciplinary Reflection on Science, Nature, Art, Human Action
and Society). Springer, Berlin-Heidelberg.

Aftanas, L., Golosheykin, S., 2005. Impact of regular meditation
practice on EEG activity at rest and during evoked negative
emotions. Int J Neurosci.115(6), 893-909.

Albert, D. Z., Loewer, B., 1988. Interpreting the many worlds
interpretation. Synthese 77, 195-213.

Albert, D. Z., 1992. Quantum mechanics and experience. Cambridge,
Mass.: Harvard Univ. Press.

Amit, D., 1989. Modeling Brain Function. Cambridge Univ. Press,
Cambridge.

Ashby, R., 1952,  Design of a brain. Chapman-Hall, London.

Aspect, A.,  Dalibard, J. and  Roger, G., 1982.
Phys. Rev. Lett., 49, 1804-1807.

Atmanspacher, H. 2007. Contextual emergence from physics to cognitive neuroscience.
J. Consciousness Studies 4 (1-2), 18-36.

Barrett,  J. A., 1999.  The quantum mechanics of minds and worlds.
Oxford Univ. Press, Oxford.

Bechtel, W., Abrahamsen,  A.,  1991. Connectionism and the mind.
Basil Blackwell, Cambridge.

Behera, L., Kar, L. and Elitzur, A. C., 2005. A recurrent quantum neural
network model to describe eye tracking of moving targets. Found. Phys. Lett.
18, N. 4, 357- 370.

Bell, J. S., 1964. Physics, 1, 195.

Bell, J. S., 1987. Speakable and unspeakable in quantum
mechanics. Cambridge Univ. Press, Cambridge.

Birkhoff, G. and von Neumann, J., 1936. The logic of quantum
mechanics. Ann. Math. 37, 823--643.

Bohm, D., Hiley,   B., 1993.  The undivided universe: an ontological
interpretation of quantum mechanics. Routledge and Kegan Paul,
London.

Brazier, M. A. B., 1970. The Electrical Activity of the Nervous
System. London: Pitman

Bredenkamp, J. (1993). Die Verknupfung verschiedener
Invarianzhypothesen im Bereich der Gedachtnispsychologie.
Zeitschrift fur Experimentelle und Angewandte Psychologie, 40,
368-385.

Busemeyer, J. B., Wang,   Z.   and Townsend,   J. T., 2006.
Quantum dynamics of human decision making, J. Math. Psychology 50,
220-241.

Busemeyer, J. B.  and Wang, Z., 2007a.  Quantum information
processing explanation for interactions between inferences and
decisions. In: Quantum interaction, Bruza,  P. D.,  Lawless,   W.,
van Rijsbergen, K. and  Sofge,  D. A.,  eds, AAAI Spring
Symposium, Technical Report SS-07-08, AAAI Press, Menlo Park, CA,
91-97.

Busemeyer, J. R., Matthew,   M. and  Wang, Z., 2007b.  A Quantum
Information Processing Theory Explanation of Disjunction Effects.
Proceedings of the Cognitive Science Society, to be published.

Buzsaki, G., 2005. Theta rhythm of navigation: link between path
integration and landmark navigation, episodic and semantic memory.
Hippocampus. 15(7), 827-40.

Clauser, J. F., Horne,  M. A., Shimony,  A. and Holt, R. A., 1982,
Phys. Rev. Lett., 49, 91.

Choustova, O. A., 2001.  Pilot wave quantum model for the stock
market, http://www.arxiv.org/abs/quant-ph/0109122.

Choustova, O., 2004. Bohmian mechanics for financial processes. J.
Modern Optics 51, n. 6/7, 1111.

Conte, E., Todarello, O., Federici,  A.,  Vitiello, T., Lopane,  M.,
Khrennikov,  A. Yu., Zbilut J. P., 2007. Some remarks on an
experiment suggesting quantum-like behavior of cognitive entities
and formulation of an abstract quantum mechanical formalism to
describe cognitive entity and its dynamics. CHAOS SOLITONS and
FRACTALS 31 (5), 1076-1088.
http://xxx.lanl.gov/abs/quant-ph/0307201.

Cossart, R., Aronov, D.  and  Yuste, R. (2003) Nature 423, 283-288.

De Baere, W., 1984.  Lett. Nuovo Cimento 39, 234-238.

Deutsch, D., 1997. The Fabric of Reality. How much can our four
deepest theories of the world explain? Publisher Allen Lane, The
Penguin Press.

Donald, M. J., 1996. On many-minds interpretation of quantum
mechanics. Preprint.

Donald, M. J., 1990. Quantum theory and brain. Proc. Royal Soc. A
427, 43-93.

Donald, M. J., 1995. A mathematical characterization of the physical
structure of observers. Found. Physics 25/4, 529-571.

Einstein, A., Podolsky, B. and Rosen, N., 1935. Phys. Rev. 47, 777-780.

Eliasmith, C., 1996.  The third contender: a critical examination of
the dynamicist theory of cognition.  Phil. Psychology 9(4), 441-463.

Fine, A., 1982.   Phys. Rev. Letters 48, 291--295.

Franco, R., 2007.   Quantum mechanics, Bayes' theorem and the
conjunction fallacy. http://www.arxiv.org/abs/quant-ph/0703222.

Geissler, H.-G., Klix, F., and Scheidereiter, U. (1978). Visual
recognition of serial structure: Evidence of a two-stage scanning
model. In E. L. J. Leeuwenberg and H. F. J. M. Buffart (Eds.),
Formal theories of perception (pp. 299-314). Chichester: John
Wiley.

Geissler, H.-G. and Puffe, M. (1982). Item recognition and no end:
Representation format and processing strategies. In H.-G. Geissler
and Petzold (Eds.), Psychophysical judgment and the process of
perception (pp. 270-281). Amsterdam: North-Holland.

Geissler, H.-G. (1983). The Inferential Basis of Classification:
From perceptual to memory code systems. Part 1: Theory. In H.-G.
Geissler, H. F. Buffart, E. L. Leeuwenberg, and V. Sarris (Eds.),
Modern issues in perception (pp. 87-105). Amsterdam: North-Holland.

Geissler, H.-G. (1985). Zeitquantenhypothese zur Struktur
ultraschneller Gedachtnisprozesse. Zeitschrift fur Psychologie, 193,
347-362. Geissler, H.-G. (1985b). Sources of seeming redundancy in
temporally quantized information processing. In G. d'Ydewalle (Ed.),
Proceedings of the XXIII International Congress of Psychology of the
I.U.Psy.S., Volume 3 (pp. 199-228). Amsterdam: North-Holland.

Geissler, H.-G. (1987). The temporal architecture of central
information processing: Evidence for a tentative time-quantum model.
Psychological Research, 49, 99-106. Geissler, H.-G. (1990).
Foundations of quantized processing. In H.-G. Geissler (Ed.),
Psychophysical explorations of mental structures (pp. 193-210).
Gottingen, Germany: Hogrefe and Huber Publishers.

Geissler, H.-G. (1992). New magic numbers in mental activity: On a
taxonomic system for critical time periods. In H.-G. Geissler, S. W.
Link, and J. T. Townsend (Eds.): Cognition, information processing
and psychophysics (pp. 293-321). Hillsdale, NJ: Erlbaum.

Geissler, H.-G. and Kompass, R. (1999). Psychophysical time units
and the band structure of brain oscillations. 15th Annual Meeting of
the International Society for Psychophysics, 7-12.

Geissler, H.-G., Schebera, F.-U., and Kompass, R. (1999).
Ultra-precise quantal timing: Evidence from simultaneity thresholds
in long-range apparent movement. Perception and Psychophysics, 6,
707-726.

Geissler, H.-G., and Kompass, R. (2001). Temporal constraints in
binding? Evidence from quantal state transitions in perception.
Visual Cognition, 8, 679-696.

Gisin, N. and Gisin, B., 1999.  Phys. Lett.  A. 260, 323.

Hameroff,  S., 1994. Quantum coherence in microtubules. A neural
basis for emergent consciousness?  J. of Consciousness Studies, 1,
91-118.

Hameroff,   S., 1998. Quantum computing in brain microtubules? The
Penrose-Hameroff Orch Or model of consciousness. Phil. Tr. Royal
Sc., London A, 1-28.

Haven, E., 2006.  Bohmian mechanics in a macroscopic quantum system.  In:
Foundations of Probability and Physics-3, Khrennikov, A. Yu.,   ed.
American Institute of Physics, 750, Melville, NY, 330-335.

Healey, R., 1984. How many worlds? Nous 18, 591-616.

Hess, K.  and Philipp, W., 2001.  Proc.
Nat. Acad. Sc., 98, 14224.

Hess, K.  and Philipp, W., 2002. Europhys. Lett. 57, 775.

Hess, K.  and Philipp, W., 2006. Bell's theorem: critique of proofs with and
without inequalities. In: Foundations of Probability and
Physics-3,  Khrennikov, A. Yu., ed.,
American Institute of Physics, 750, Melville, NY, 150-157.

Hiley,  B.,  Pylkk\"anen, P., 1997.  Active information and
cognitive science -- A reply to Kiesepp\"a. In: Brain, mind and
physics.  Editors: Pylkk\"anen, P., Pylkk\"o, P., Hautam\"aki, A.
IOS Press, Amsterdam.

Hiley, B., 2000. Non-commutavive geometry, the Bohm interpretation
and the mind-matter relationship. Proc. CASYS 2000, Liege, Belgium.

Hopfield,  J. J., 1982. Neural networks and physical systems with
emergent collective computational abilities. Proc. Natl. Acad. Sci.
USA 79,  1554-2558.

Hoppensteadt, F. C., 1997.  An introduction to the mathematics of
neurons: modeling in the frequency domain.  Cambridge Univ. Press,
New York.

Ikegaya, Y., Aaron, G., Cossart, R., Aronov, D., Lampl, I., Ferster,
D. and Yuste, R. (2004) Science 304, 559-564.

Jibu, M., Yasue,  K., 1992. A physical picture of Umezawa's quantum
brain dynamics. In  Cybernetics and Systems Research, ed. R. Trappl,
World Sc., London.

Jibu, M.,  Yasue,  K., 1994.  Quantum brain dynamics and
consciousness. J. Benjamins Publ. Company, Amsterdam/Philadelphia.

Kerr, J. N., Greenberg, D. and Helmchen, F. (2005) Proc. Natl. Acad.
Sci. U. S. A 102, 14063-14068.

Khrennikov, A. Yu., 1999a. Classical and quantum mechanics on
information spaces with applications to cognitive, psychological,
social and anomalous phenomena. Found. Phys. 29,  1065-1098.

Khrennikov, A. Yu., 1999b. Interpretations of Probability, VSP Int.
Sc. Publishers, Utrecht/Tokyo, 1999 (second edition, 2004).

Khrennikov,  A. Yu., 2000.  Classical and quantum mechanics on
$p$-adic trees of ideas. BioSystems   56, 95-120.

Khrennikov, A. Yu. (editor), 2001.  Foundations of
Probability and Physics, Quantum Probability and White Noise
Analysis 13, WSP, Singapore.

Khrennikov, A. Yu. (editor), 2002a. Quantum
Theory: Reconsideration of Foundations, Ser. Math. Modeling
2, V\"axj\"o Univ. Press, V\"axj\"o.

Khrennikov, A. Yu., 2002b. On cognitive experiments to test
quantum-like behaviour of mind. Rep. V\"axj\"o Univ.: Math. Nat. Sc.
Tech., N 7; http://xxx.lanl.gov/abs/quant-ph/0205092.

Khrennikov, A. Yu. (editor), 2003a.
Foundations of Probability and Physics-2, Ser. Math. Modeling
5, V\"axj\"o Univ. Press,  V\"axj\"o, 2003.

Khrennikov,  A. Yu., 2003b. Quantum-like formalism for cognitive
measurements. Biosystems 70, 211-233.

Khrennikov, A. Yu., 2004a. Information Dynamics in Cognitive,
Psychological, Social, and Anomalous Phenomena (Fundamental Theories
of Physics). Springer, Berlin-Heidelberg.

Khrennikov, A. Yu., 2004b. Probabilistic pathway representation of cognitive information.
J. Theor. Biology, 231, 597-613.

Khrennikov, A. Yu., 2005a. A pre-quantum classical statistical model
with infinite-dimensional phase space. J. of Physics A, Math. and
General 38 (41), 9051-9073.

Khrennikov, A. Yu., 2005b. Generalizations of quantum mechanics
induced by classical statistical field theory. Foundations of
Physics Letters  18 (7),  637-650.

Khrennikov, A. Yu.  and Volovich, I. V., 2005.
Soft Computing 10,  521 - 529.

Khrennikov, A. Yu., 2006a. Quantum-like brain: "Interference of
minds" Biosystems 84 (3),  225-241.

Khrennikov, A. Yu., 2006b. On the problem of hidden variables for
quantum field theory. Nuovo Cimento B  121 (5), 505-521.

Khrennikov, A. Yu., 2006c. Nonlinear Schrodinger equations from
subquantum classical statistical field theory. Physics Letters  A 357
(3), 171-176,

Khrennikov, A. Yu., 2006d. To quantum mechanics through random
fluctuations at the Planck time scale.
http://www.arxiv.org/abs/hep-th/0604011; Nuovo Cimento (accepted
for publication).

Klix, F., and van der Meer, E. (1978). Analogical reasoning - an
approach to mechanisms underlying human intelligence performances.
In F. Klix (Ed.), Human and artificial Intelligence (p. 212).
Berlin: Deutscher Verlag der Wissenschaften.

Klyshko, D. N., 1993a. Phys. Lett. A  172,  399-403.

Klyshko, D. N., 1993b. Phys. Lett. A 176, 415-420.

Klyshko, D. N., 1995. Annals of New York Academy of Science 755,  13-27.

Klyshko, D. N., 1996a. Phys. Lett. A  218, 119-12.

Klyshko, D. N., 1996b. Laser Physics 6,  1056-1076.

Klyshko, D. N., 1998a. Phys. Lett. A  247,  261-266.

Klyshko, D. N., 1998b.  Uspehi Fizicheskih Nauk 168,
975-1015.

Kristofferson, M. W. (1972). Effects of practice on
character-classification performance. Canadian Journal of
Psychology, 26, 540-560.

Kristofferson, A. B. (1980). A quantal step function in duration
discrimination. Perception and Psychophysics, 27, 300-306.

Kristofferson, A. B. (1990). Timing mechanisms and the threshold for
duration. In Geissler, H.-G. (Ed., in collaboration with M. H.
Muller  and  W. Prinz), Psychophysical explorations of mental
structures (pp. 269-277). Toronto: Hogrefe  and  Huber Publishers.

Larsson, J.-A., 2000. Quantum Paradoxes, Probability Theory, and Change of
Ensemble, Link\"oping Univ. Press, Link\"oping.

Lockwood, M., 1989. Mind, Brain and Quantum. Oxford, Blackwell.

Lockwood, M., 1996. Many minds interpretations of quantum mechanics.
British J. for the Philosophy of Sc. 47/2, 159-88.

Loewer, B., 1996. Comment on Lockwood. British J. for the Philosophy
of Sc. 47/2, 229-232.

Luczak, A., Bartho, P., Marguet, S. L., Buzsaki, G., Hariis, K.D,
2007. Neocortical spontaneous activity in vivo: cellular
heterogeneity and sequential structure. Preprint of CMBN, Rutgers
University.

MacLean, J. N., Watson, B. O., Aaron, G. B. and Yuste, R. (2005)
Neuron 48, 811-823.

Mao, B. Q., Hamzei-Sichani, F., Aronov, D., Froemke, R. C. and
Yuste, R. (2001). Neuron 32, 883-898.

N\"a\"at\"anen, R.,  (1992). Attention and brain function.
Lawrence Erlbaum Ass. Publ., Hillsdale, NJ.

Orlov, Y. F., 1982.  The wave logic of consciousness: A hypothesis.
Int. J. Theor. Phys. 21, N 1, 37-53.

Pearle, P., 1970.  Phys. Rev. D 2,  1418.

Penrose, R., 1989. The emperor's new mind. Oxford Univ. Press,
New-York.

Penrose, R., 1994.  Shadows of the mind. Oxford Univ. Press, Oxford.

Penrose, R., 2005. The Road to Reality : A Complete Guide to the
Laws of the Universe. Knopf Publ.

Pitowsky, I., (1982).  Phys. Rev. Lett 48, N.10, 1299-1302.

Plotnitsky, A., 2001. Reading Bohr: Complementarity, Epistemology,
Entanglement, and Decoherence. Proc. NATO Workshop Decoherence and
its Implications for Quantum Computations, Eds. A.Gonis and
P.Turchi, p.3--37, IOS Press, Amsterdam.

Plotnitsky, A., 2002. Quantum atomicity and quantum information:
Bohr, Heisenberg, and quantum mechanics as an information theory,
Proc. Conf.  Quantum theory: reconsideration of foundations, ed: A.
Yu. Khrennikov, Ser. Math. Modelling
 2, 309-343, V\"axj\"o Univ. Press,  V\"axj\"o.

Plotnitsky, A., 2007. Reading Bohr: Physics and Philosophy
(Fundamental Theories of Physics). Springer, Berlin-Heidelberg.

Pylkk\"anen, P., 2006. Mind, Matter and the Implicate Order (The
Frontiers Collection). Springer, Berlin-Heidelberg.

Puffe, M. (1990). Quantized speed-capacity relations in short-term
memory. In H.-G. Geissler (Ed., in collaboration with M. H. Muller
and W. Prinz), Psychophysical exploration of mental structures (pp.
290-302). Toronto: Hogrefe and Huber Publishers.

Sanchez-Vives, M. V. and McCormick, D. A. (2000) Nat. Neurosci. 3,
1027-1034.

Schr\"odinger, E., 1982. Philosophy and the Birth of Quantum
Mechanics. Edited by M. Bitbol, O. Darrigol. Editions Frontieres,
Gif-sur-Yvette.

Shu, Y. S., Hasenstaub, A. and McCormick, D. A. (2003) Nature 423, 288-293.

Stapp, H. P.,  1993. Mind, matter and quantum mechanics.
Springer-Verlag, Berlin-New York-Heidelberg.

Strogatz,  S. H., 1994.  Nonlinear dynamics and chaos with
applications to physics, biology, chemistry, and engineering.
Addison Wesley, Reading, Mass.

Svozil, K., 2006. Quantum Logic (Discrete Mathematics and
Theoretical Computer Science). Springer, Berlin-Heidelberg.

Teghtsoonian, R. (1971): On the exponents in Stevens' law and on the
constant in Ekman's law. Psychological Review, 78, 71 - 80.

van Gelder, T., Port, R., 1995. It's about time: Overview of the
dynamical approach to cognition. in  Mind as motion: Explorations in
the dynamics of cognition. Ed.: T. van Gelder, R. Port. MITP,
Cambridge, Mass, 1-43.

van Gelder, T., 1995. What might cognition be, if not computation?
J. of Philosophy 91, 345-381.

Vitiello, G., 2001.  My double unveiled - the dissipative quantum
model of brain. J. Benjamins Publ. Company, Amsterdam/Philadelphia.

von Bekesy, G. (1936). Uber die Horschwelle und Fuhlgrenze langsamer
sinusformiger Luftdruckschwankungen. Annalen der Physik, 26,
554-556.

Volovich, I., 2001, Quantum cryptography in space and Bell's theorem. In:
Foundations of probability and physics,
Khrennikov, A. Yu.,  ed.,   Quantum Prob. White Noise Anal. 13,
World Sci. Publ., River Edge, NJ, pp. 364--372.

von Neumann, J.,  1955. Mathematical foundations of quantum
mechanics. Princeton Univ. Press, Princeton, N.J..

Weihs, G.,  Jennewein, T.,  Simon, C.,  Weinfurter, H., and Zeilinger, A., 1998.
Phys. Rev. Lett., 81,  5039.

Weihs, G., 2007.  A test of Bell's inequality with spacelike separation.
In:  Foundations of Probability and
Physics-4, Adenier, G., Fuchs,  C.  and Khrennikov, A. Yu.,  eds.,
American Institute of Physics,  889, Melville, NY, pp. 250-262.

Whitehead, A. N., 1929. Process and Reality: An Essay in Cosmology.
Macmillan Publishing Company, New York.

Whitehead, A. N., 1933.  Adventures of Ideas.  Cambridge Univ.
Press, London.

Whitehead, A. N., 1939. Science in the modern world. Penguin,
London.

\end{document}